\documentclass[aps,pra,floatfix,onecolumn,tightenlines,superscriptaddress,notitlepage]{revtex4-1}

\usepackage{graphicx}
\usepackage{amsmath}
\usepackage{latexsym}
\usepackage{graphicx}
\usepackage{multirow}
\usepackage{url}
\usepackage[bookmarks, colorlinks=true, pdftitle={Testing quantum physics in space using high-mass matter-wave interferometry},
            pdftex
            ]{hyperref}
                        
\makeatother

\begin{document}

\title{Testing quantum physics in space using high-mass matter-wave interferometry}

\author{Kaltenbaek, Rainer}
\email[E-Mail: ]{rainer.kaltenbaek@univie.ac.at}
\affiliation{Vienna Center for Quantum Science and Technology, Faculty of Physics, University of Vienna, Boltzmanngasse 5, 1090 Vienna, Austria}

\begin{abstract}
Quantum superposition is central to quantum theory but challenges our concepts
of reality and spacetime when applied to macroscopic objects like Schr\"odinger's cat. For that reason, it has been a long-standing question whether quantum physics remains valid unmodified even for truly macroscopic objects. By now, the predictions of quantum theory have been confirmed via matter-wave interferometry for massive objects up to 
$10^4\,$ atomic mass units (amu). The rapid development of new technologies promises to soon allow tests of quantum theory for significantly higher test masses by using novel techniques of quantum optomechanics and high-mass matter-wave interferometry. Such experiments may yield novel insights into the foundations of quantum theory, pose stringent 
limits on alternative theoretical models or even uncover deviations from quantum physics. However, performing experiments of this type on Earth may soon face
principal limitations due to requirements of long times of flight, ultra-low vibrations, and extremely high vacuum. Here, we present a short overview of recent developments towards the implementation of the proposed space-mission MAQRO, which promises to overcome those limitations and to perform matter-wave interferometry in a parameter regime orders of magnitude beyond state-of-the-art.
\end{abstract}

\maketitle

\section{Introduction}
Some of the central concepts of quantum physics have been a topic of discussion from the start. In particular, quantum superposition in the case of macroscopic objects like Schr\"odinger’s cat\cite{Schroedinger1935a} challenges our concepts of reality. For massive particles with a mass up to $10^4$ atomic mass units (amu), the predictions of quantum theory have been confirmed experimentally\cite{Eibenberger2013a}. While this is still far from the mass of Schr\"odinger's cat, experiments may soon be able to prepare quantum superpositions of objects visible to the naked eye. Various alternative theoretical models (collapse models) predict observable deviations from quantum theory in that context\cite{Adler2009a,Bassi2013a}. Independent of whether any of those models are correct or not, achieving quantum control over sufficiently macroscopic physical systems will mark a milestone towards systematically exploring an entirely new parameter regime. Eventually, high-mass matter-wave interferometry may allow testing deviations from quantum theory due to metric fluctuations due to quantum gravity\cite{Breuer2009a} or gravitational-wave background\cite{Jaekel1994a,Lamine2006a} or, in the presence of a gravitational field, decoherence due to time dilation\cite{Pikovski2015a}. Recently, it has been suggested that high-mass matter-wave interferometry may even be sensitive to certain types of dark matter\cite{Riedel2013a,Bateman2015a}. 

\section{Limitations in ground-based experiments}
To enable matter-wave interferometry for masses beyond current experiments, novel techniques are being developed, like optical time-domain ionizing matter-wave (OTIMA) inter\-fero\-metry\cite{Nimmrichter2011a,Nimmrichter2011b} or using optically trapped particles for far-field\cite{RomeroIsart2011b} or near-field interferometry\cite{Bateman2014a}. These approaches may allow achieving tests of quantum theory for test masses up to $10^6\,$amu or even $10^8\,$amu over the next years. Beyond that, ground-based experiments may efforts may soon face principal limitations due to limited free-fall times as well as limited quality of vacuum and micro-gravity\cite{Kaltenbaek2012b,Kaltenbaek2015a}. Given this mass limit, ground-based experiments may eventually allow decisive tests\cite{Nimmrichter2011b,RomeroIsart2011b} of the continuous spontaneous localization (CSL) model\cite{Ghirardi1990a} and the quantum-gravity (QG) model of Ellis et al\cite{Ellis1989a}. Testing quantum physics for higher masses, and testing other collapse models like that of K\'arolyh\'azy\cite{Karolyhazy1966a} or Di\'osi-Penrose\cite{Diosi2008a} seems to be beyond ground-based experiments. The same holds true for more ambitious tests of metric fluctuations, time-dilation or dark matter.

It is conceivable that some of those limitations can be overcome by using magnetic levitation of superconducting spheres\cite{RomeroIsart2012a}. Still, this approach has to be investigated in more detail to assure its applicability for testing quantum physics under realistic conditions (i.e., in the presence of field fluctuations, vibrations, material inhomogeneities, etc.).

Given the rapid development of space technology, e.g., in the context of LISA Pathfinder\cite{Armano2009}, using a space environment for quantum experiments is becoming an attractive alternative. For this reason, in 2010, we proposed a medium-sized space mission MAQRO to perform high-mass matter-wave interferometry in space\cite{Kaltenbaek2012b}. Here, we present an short overview of the MAQRO mission proposal and its current status.

\section{The MAQRO mission proposal}
The goal of MAQRO is to perform decisive tests of quantum physics by optimally harnessing the unique opportunities offered by a space platform, i.e., microgravity, and the possible low temperatures and ultra-high vacuum outside the spacecraft. In particular, MAQRO will take full advantage of the rich heritage of several missions and mission proposals. It will use the same spacecraft, carrier and orbit as LISA Pathfinder (LPF)\cite{Armano2009}, microthrusters as used in GAIA, LPF and Microscope\cite{Liorzou2014a}, and Onera inertial sensors based on established technology used in missions like Microscope and GOCE\cite{Marque2010a}. The experiments performed in MAQRO will be matter-wave interferometry with high-mass test particles: dielectric nanospheres of different radii and materials to quantitatively test quantum physics over a wide range of parameters. Compared to the present mass record of $10^4\,$amu\cite{Eibenberger2013a}, MAQRO aims at testing quantum physics with test masses up to several $10^{10}\,$amu.

Achieving this goal and the corresponding requirements of $\sim 100\,$s coherence time will require extremely low vacuum levels of $\sim 10^{13}\,$Pa and temperatures of $\lesssim 20\,$K for the environment and $\lesssim 25\,$K\cite{Kaltenbaek2015a}. In the original proposal of MAQRO\cite{Kaltenbaek2012b}, we suggested to achieve the ultra-high vacuum level and low environment temperature by using a platform outside the spacecraft.

\subsection{Mission configuration}
The central component of MAQRO is an optical bench mounted outside the spacecraft and isolated from the hot spacecraft via a structure of three thermal shields\cite{Kaltenbaek2012b}. The design of the heat-shield structure and the optical bench was optimized in two thermal studies\cite{Hechenblaikner2014a,Pilan-Zanoni2015a}. In particular, we showed that the vacuum achievable on the optical bench should only be limited by interplanetary vacuum in a Lissajous orbit around the Earth-Sun Lagrange point L1 (or L2)\cite{Kaltenbaek2012b}, compatible with the requirements of MAQRO\cite{Kaltenbaek2015a}. Our thermal studies showed that the environment temperature achievable via passive cooling is $\sim 25\,$K for the optical bench and down to $\sim 12\,$K for a small test volume around the experimental region -- the ``test volume''. This is achieved by placing only the absolute minimum of optical components and dissipative elements on the optical bench, and by optimizing the coating of optical elements and the optical bench. Using reflective instead of refractive optics for the on-bench imaging system allowed reducing the temperature of the test volume from $\sim 16\,$K to $\sim 12\,$K.

The thermal-shield structure will be mounted outside the spacecraft and always pointing to deep space with the spacecraft in a sun-synchronous orbit around L1 (or L2). The orbit was chosen for several reasons: (1) the high thermal stability achievable\cite{Kaltenbaek2012b}, (2) the ultra-high interplanetary vacuum, (3) the low temperature achievable via passive cooling, (4) the low gravitational field gradients, and (5) the technological heritage of LPF.

We choose a nominal mission life-time of two years with possible extension in order to allow for the accumulation of a sufficient amount of data to achieve the scientific goals. The spacecraft, launcher and orbit were chosen identical to LPF apart from larger fuel tanks to accommodate the longer mission lifetime.

\subsection{The experiment}
In contrast to the original MAQRO proposal\cite{Kaltenbaek2012b}, which was based on double-slit-type far-field matter-wave interference using a novel form of quantum state preparation, the updated mission proposal\cite{Kaltenbaek2015a}, takes advantage of established matter-wave-interferometry techniques to perform near-field interferometry\cite{Bateman2014a}.

The central approach remains the same: (1) optically trap a dielectric particle, (2) cool its center-of-mass motion close to the quantum ground-state using optomechanical techniques, (3) release the particle and let the wavefunction expand for a time $t_1$, (4) prepare a non-classical state of motion of the particle, (5) let the wavefunction freely evolve for a time $t_2$, (5) measure the position of the test particle. This procedure is then repeated many times to gather enough statistics to determine the interference visibility. In order to test the predictions of quantum theory, such experimental runs are repeated for different materials and different particle sizes. 

In contrast to the original proposal, we have $t_1 \approx t_2$, and step (4) is the application of a phase grating of UV light with a wavelength of $\sim 200\,$nm. Central advantages of the new approach are that the total measurement time $t_1 + t_2$ is shorter, that the experiment relies on laser wavelengths that are already available in space ($1064\,$nm) or may be soon ($\sim 200\,$nm CW on the mW level) with a manageable amount of delta development, and that the matter-wave interference visibility to be expected can be very high.

MAQRO will use a combination of ion and optical trapping to provide a reliable source of high-mass test particles for the experiment. They will be guided via hollow-core fibre from inside the spacecraft to the optical bench outside. Buffer gas in the hollow-core fibres will ensure that the temperature of the test particles will only be slightly above the environment temperature. Before loading the test particle into the experimental region, we plan to discharge using UV radiation\cite{Kaltenbaek2015a}. We will also aim at minimizing the time of optical trapping of the particle in order to keep it cool. Moreover, a different particle will be used for each experimental run.

\section{Outlook}
In the near future, we will further improve and detail the MAQRO mission design, perform preliminary experiments on ground, extend and intensify our our international collaboration for realizing MAQRO, and we will prepare for the submission of an improved mission proposal for the next mission opportunity.

\section{Conclusions}
We presented a short overview of the current status of the proposal MAQRO of a medium-sized space mission for testing the foundations of quantum physics -- its goals, the mission outline and the next steps towards implementing that mission.

\section*{Acknowledgements}
We acknowledge support from the FFG Austrian Research Promotion Agency (MAQROsteps, project no. 3589434).

\section*{References}

\bibliographystyle{apsrev4-1}
\bibliography{/home/rainer/physik/rk}  

\end{document}